\documentclass[%
 aip,
 amsmath,amssymb,
 reprint,%
]{revtex4-1}

\usepackage{graphicx}
\usepackage{dcolumn}
\usepackage{bm}
\usepackage{ulem}
\usepackage{hyperref}
\usepackage{color}

\usepackage[utf8]{inputenc}
\usepackage[T1]{fontenc}
\usepackage{mathptmx}

\begin{document}

\preprint{AIP/123-QED}

\title{Planar Hall effect caused by the memory of antiferromagnetic domain walls in Mn$_3$Ge}

\author{Liangcai Xu$^{1,2}$, Xiaokang Li$^{1}$, Linchao Ding$^{1}$,  Kamran Behnia$^{2}$, Zengwei Zhu$^{*,}$}

 \affiliation{
	(1) Wuhan National High Magnetic Field Center and school of physics, Huazhong University of Science and 		Technology, Wuhan 430074,China\\
	(2) Laboratoire de Physique et d'Etude des Mat\'{e}riaux (CNRS), ESPCI Paris, PSL Research University, 75005 Paris, France}

\date{\today}

\begin{abstract}

In Mn$_3$X (X=Sn, Ge) antiferromagnets domain walls are thick and remarkably complex because of the non-collinear arrangement of spins in each domain. A planar Hall effect (PHE), an electric field perpendicular to the applied current but parallel to the applied magnetic field, was recently observed inside the hysteresis loop of Mn$_3$Sn. The sign of the PHE displayed a memory tuned by the prior orientation of the magnetic field and its history. We present a study of PHE in Mn$_3$Ge extended from room temperature down to 2 K and show that this memory effect can be manipulated by either magnetic field or thermal cycling. We show that the memory can be wiped out if the prior magnetic field exceeds 0.8 T or when the temperature exceeds $T_\mathrm{N}$. We also find a detectable difference between the amplitude of PHE with zero-field and field thermal cycling. The ratio between the PHE and the anomalous Hall effect (AHE) decreases slightly as temperature is increased from 2 K to $T_{\rm{N}}$, tracks the temperature dependence of magnetization. This erasable  memory effect may be used for data storage.
\end{abstract}

\maketitle

The discovery of a large room-temperature anomalous Hall effect(AHE)  in non-collinear antiferromagnetic Mn$_3$X (X = Sn, Ge)~\cite{Nakatsuji2015,Nayak2016,Kiyohara2016} was theoretically anticipated~\cite{Chen2014,Kubler_2014}. The non-trivial Berry curvature of the electron wave-function~\cite{Xiao2010,Nagaosa2010} is widely believed to be the origin of this phenomenon. It stimulated numerous experimental studies and was followed by the discovery of other topological magnets~\cite{Yin2020,Morali2019,Liu2019,Belopolski2019,Ding2019}.

Other  counterparts of the AHE were  observed in the Mn$_3$X family. the list includes the anomalous Nernst effect~\cite{Li2017,Ikhlas2017,Xu2020}, the anomalous thermal Hall effect~\cite{Li2017,Sugii2019,Xu2020}, a large Kerr effect~\cite{Higo2018} as well as  high-frequency AHE~\cite{Matsuda2020}.The Mn$_3$X family provide also candidates for  antiferromagnetic spintronic devices, which are attractive due to the absence of perturbing stray field and their fast response~\cite{Duine2006,Gomonay2014,review2018}. There is a prospect for high-speed operation of the devices. They can reach THz range, which is three orders of magnitude faster than those made of ferromagnets\cite{Jungwirth2018}. Thus, the applications of the devices may extend to the areas like terahertz information technologies or artificial neural networks\cite{Jungwirth2018}. Non-collinear antiferromagnets host the above-mentioned AHE, and possibly a topological Hall effect due to the chirality of their spin texture\cite{Fischer2014,Li2018}. This could allow the generation of spin polarized charge current\cite{Jakub2017} and other applications.


Theoretical studies have associated the magnetic texture with cluster octupoles\cite{Suzuki2017} and found that its topological defects, including the walls separating magnetic domains have non-trivial properties~\cite{Liu2017}. The antiferromagnetic structure of Mn$_3$X can be characterized by the cluster octupole moments for the $D_{6h}$ irreducible representations.\cite{Suzuki2017} These domain walls were scrutinized in a number of experiments\cite{Higo2018,Li2018Domain,Li2019}. In particular, in presence of domain walls,  a topological Hall effect, presumably arising from the real-space Berry curvature, was observed  first in Mn$_3$Sn crystals~\cite{Li2018Domain} and subsequently in thin films~\cite{Taylor2020}.

The subject of the present paper is the planar Hall effect (PHE). Such a signal (i.e. an electric field, which resides in the same plane with magnetic field and electric current, but is normal to the latter) was recently reported in Mn$_3$Sn~\cite{Li2019}.  Together with its thermoelectric counterpart (i.e. a planar Nernst effect) and off-diagonal magnetization, the PHE emerges in a narrow magnetic field where domains of opposite polarity co-exist~\cite{Li2019}. Its sign in Mn$_3$Sn was found to have a memory depending on the rotational history of magnetic field in respect with the specific chirality of  domain walls. This  constitutes a type of memory formation for recording a direction~\cite{Keim2019}. The previous study on Mn$_{3}$Sn~\cite{Li2019} was restricted to room temperature. Here, we present a study of the PHE and the memory effect in Mn$_3$Ge in a wide temperature range. In contrast to Mn$_{3}$Sn, where the triangular spin order is destroyed below  $\sim$50 K\cite{Nakatsuji2015,Ohmori1987}, in Mn$_{3}$Ge one can follow the AHE and  associated phenomena  down to 2 K ~\cite{Nayak2016,Kiyohara2016} and even below 1 K ~\cite{Xu2020}. Moreover, the lower N\'eel temperature ($T_{\rm{N}}=$ 370 K\cite{Xu2020}) makes it possible to go above the ordering temperature by sweeping to 400 K.  We find that both the PHE and its memory exist in Mn$_3$Ge. At room temperature, the PHE/AHE ratio  is somewhat smaller than what was found in Mn$_3$Sn, presumably due to a difference in the width of the hysteresis loop and coercive field scales.  The PHE/AHE ratio presents a temperature dependence, which roughly mirrors the temperature dependence of magnetization. By performing a thermal cycle across $T_{\rm{N}}$, we show that the memory can be erased by warming the sample above its ordering temperature. We also quantified the threshold of erasure for prior magnetic field required to wipe out the PHE~\cite{Li2019}.

\begin{figure}[htb]
\includegraphics[width=8.5cm]{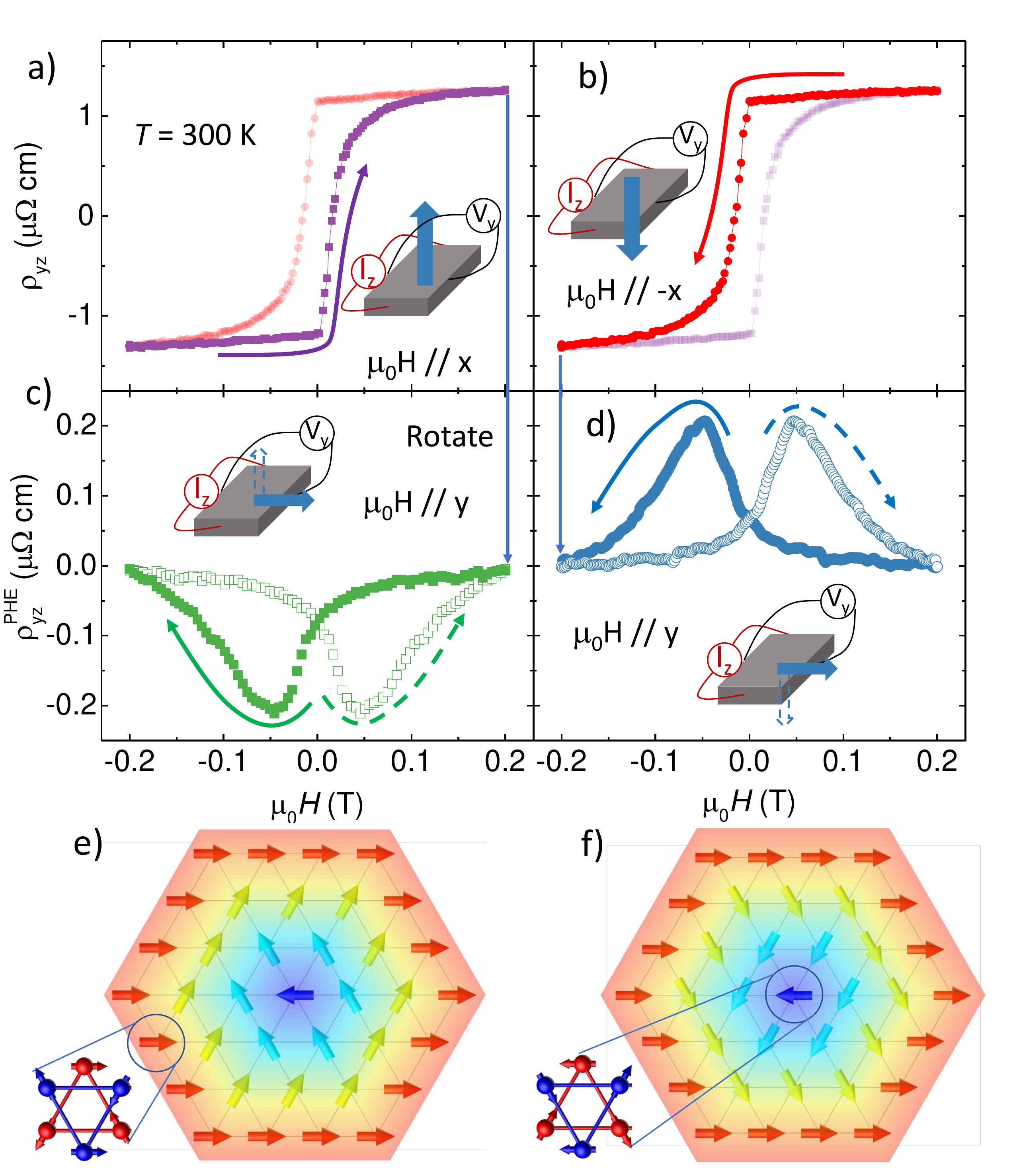}
\caption{\textbf{Anomalous Hall effect and planar Hall effect at room temperature}
a) and b) Anomalous Hall effect (AHE) with field sweeping from $\mathrm{\mu_{0}H // -x}$ $\rightarrow$ $\mathrm{\mu_{0}H // x}$  and $\mathrm{x}$ $\rightarrow$ $\mathrm{-x}$. Curves with transparent symbols refer to the opposite sweeping direction. c)  Planar Hall effect (PHE)  with an prior field rotated from the final state of a). d)  PHE with an initial field  rotated from the final state of b).  Note that the magnetic field has been rotated from out-of-plane to in-plane to induce chirality of the antiferromagnetic domain walls. The sign of PHE is set by  the  history of prior orientation before rotation. e) and f) show two  domain wall chiralities~\cite{Li2019}. Each arrow represents an octupole moment~\cite{Suzuki2017,Sugimoto2020}.
\label{fig:Ratio}}
\end{figure}

\begin{figure}[htb]
\includegraphics[width=8.5cm]{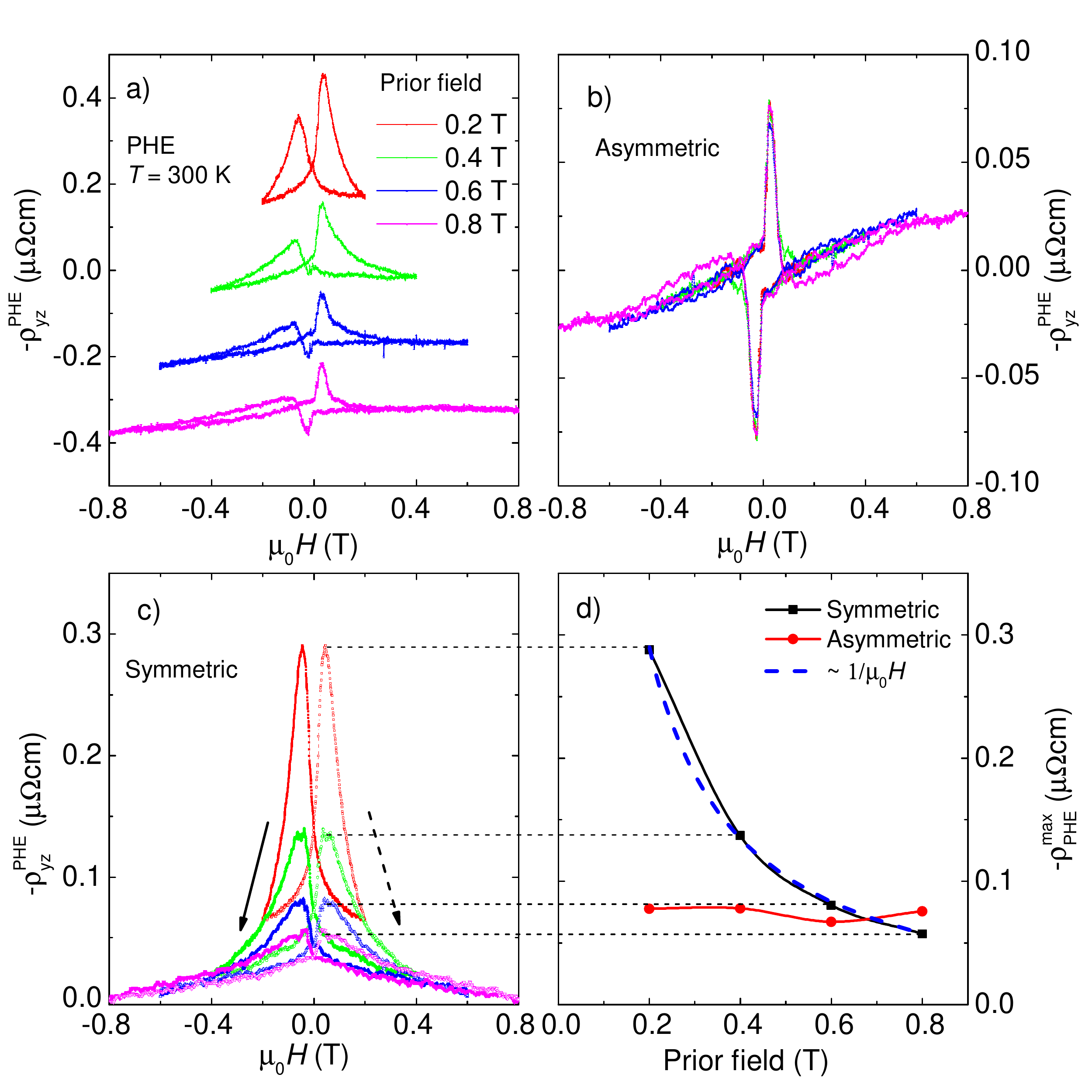}
\caption{\textbf{Dependence of the planar Hall effect on the amplitude of the prior field:} (a) Raw PHE data  after rotating the field from $+\pi/2$ to $\pi$. (b) The asymmetric component of the measured signal. (c) The symmetric component. d) The peak amplitude of the two components of PHE as a function of the amplitude of the prior magnetic field. The symmetric part follows $1/\mu_{0}H$, while the asymmetric part is almost flat, indicating that is a residual AHE signal.\label{fig:field dependent}}
\end{figure}

\begin{figure}[htb]
\includegraphics[width=8.5cm]{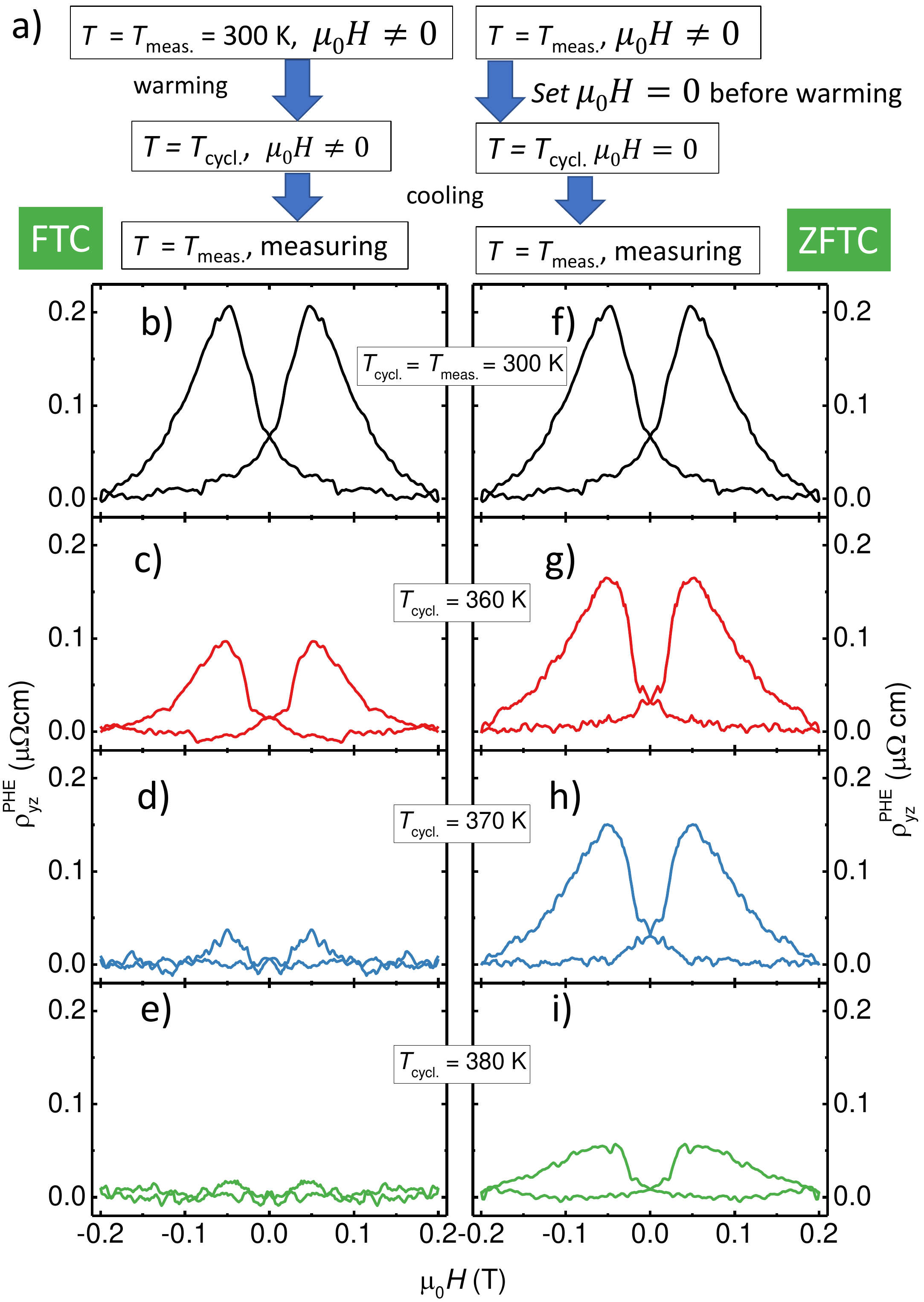}
\caption{\textbf{The evolution of  planar Hall effect with thermal cycling} a) Two distinct procedures. In field-thermal-cycling(FTC), the sample was warmed from $T_\mathrm{meas.}$ to cycling temperature($T_\mathrm{cycl.}$) and cooled down to $T_\mathrm{meas.}$ and then measurement was performed. In the zero-field-thermal-cycling(ZFTC), the warm-up to $T_\mathrm{cycl.}$ and the cool-down to $T_\mathrm{meas.}$ were performed in absence of magnetic field. b,c,d,e) show the FTC planar Hall effect with different cycling temperatures. Panel f.g.h.i) show the ZFTC residual planar Hall effect. All  measurements were performed at $T_\mathrm{meas.}=300 K$. In both cases, the planar Hall effect gradually decreases with the approach to $T_\mathrm{N}$. But it drops faster in the FTC process than in the ZFTC process.
\label{fig:evolution}}
\end{figure}

\begin{figure}[htb]
\includegraphics[width=8.5cm]{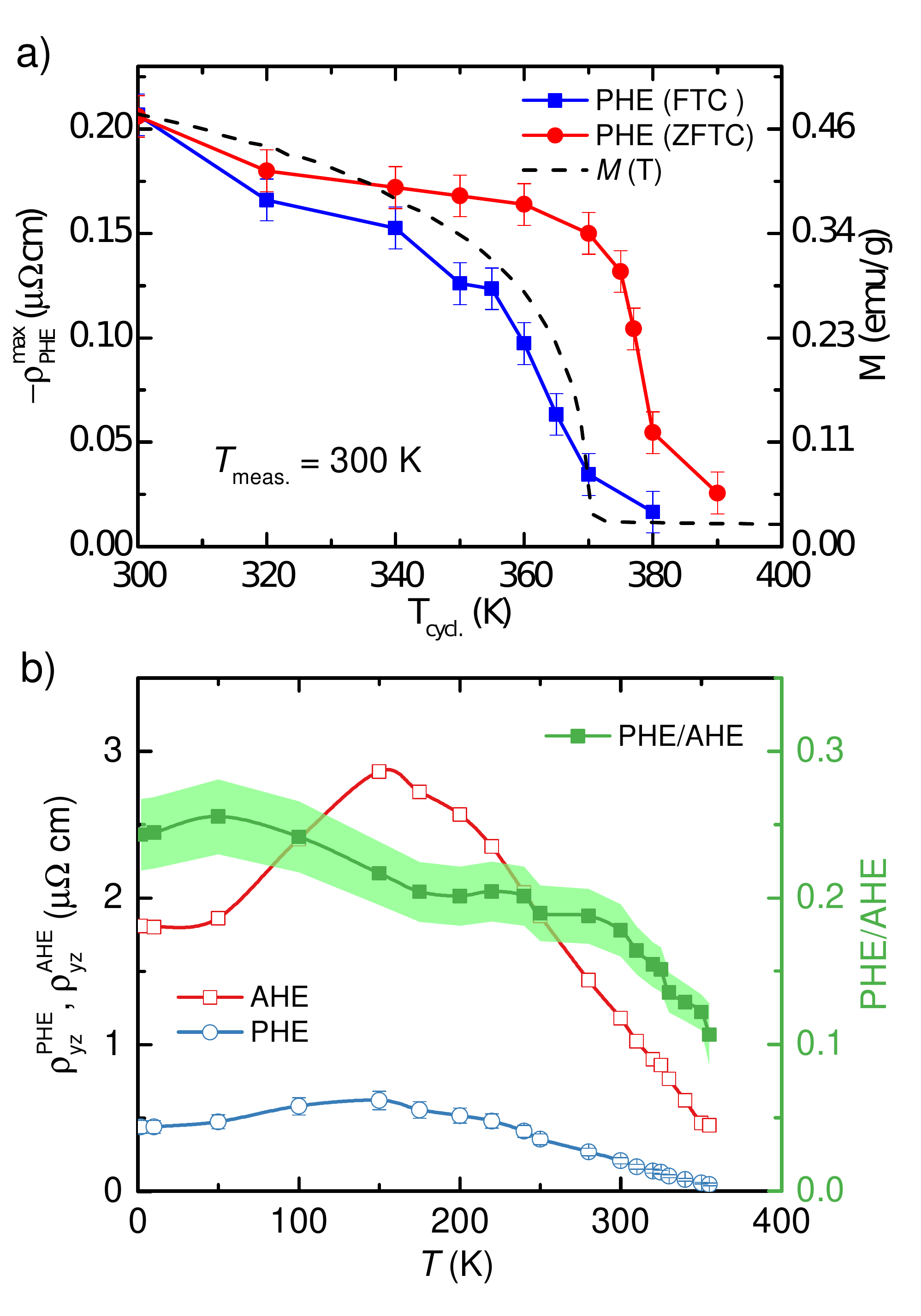}
\caption{\textbf{Evolution with temperature:} a) The blue and the red curves represent the evloution of the FTC and ZFTC PHE as a function of cycling temperature. The PHE in the ZFTC process survives at temperatures slightly exceeding the N\'{e}el temperature, as observed by the $M$(T) curve (dashed line). b)  The temperature dependence of the AHE (red), PHE (blue) and their ratio (green).
\label{fig:compare}}
\end{figure}

The single crystals were grown by Bridgman-Stockbarger technique~\cite{Xu2020}. They were subsequently cut into approximate dimensions of $1\times2\times0.5\rm{mm}^3$ with a wire saw.

Fig.\ref{fig:Ratio} a) and c)) shows the measurement procedure to obtain the AHE and PHE. The electric current was applied along $z$[0001] and the voltage was measured along $y$[01$\bar{1}$0]. The AHE was recorded by orienting the magnetic field along $x$[2$\bar{1}\bar{1}$0], and sweeping it from negative to positive values(Fig.\ref{fig:Ratio} a)) and/or vice versa(Fig.\ref{fig:Ratio} b)).  As seen in the Fig.\ref{fig:Ratio} a), three different regimes (I, II, III) replace each other as the field is swept. In regime I, there  is a single domain induced by magnetic field. Above a threshold field $\mu_0 H_0$, new domains with opposite polarity nucleate and regime II starts. At sufficiently large field, the  whole system becomes single domain but with spins oriented in opposite orientation compared to regime I. The amplitude and field dependence of  $\rho_{yz} $ is almost similar to what was  previously reported~\cite{Nayak2016,Kiyohara2016}.

To measure PHE, the magnetic field was rotated from $x$ to $y$ (By PHE, we mean the symmetric part of the measured signal). Then, the magnetic field (kept oriented along $y$) was swept from negative to positive values, using the same electrodes for current (along $z$) and  voltage (along $y$). The measured signal was therefore a version of the so-called planar Hall effect, because the three vectors (the electric current, the electric field and the magnetic field) were residing in the same plane. To assure high accuracy PHE signal, the tilted angle between the magnetic field and the current-electric field was kept less than 2 degrees. We define the magnetic field at which the rotation took place as the prior magnetic field ($\mu_0H_p$).

Since the PHE is very sensitive to the field orientation, the field angle is to be carefully tuned. The measured signal has two components which are even and odd as a function of magnetic field. The asymmetric part is the residual AHE arising from imperfect alignment, but the symmetric part (even in magnetic field) can only be produced by a genuine PHE~\cite{Li2019}.  This  signal  emerges only in regime II. In this field window,  the two domains with opposite polarities compensate the transverse response of each other. The measured electric field is produced by the domain wall whose spin configuration is perpendicular to the two domains sandwiching it (See Fig.\ref{fig:Ratio} e, f). The sign of this electric field is set by the chirality of the domain wall. As seen in Fig.\ref{fig:Ratio} c) and d), two distinct types of PHE curves similar in amplitude but opposite in sign can be obtained. What controls the sign is the history of the magnetic field prior to  sweeping field.  The prior field orientation determines the sign of the PHE. The domain wall separating two domains with 0 and $\pi$ (the angle is defined as between the x-axis and a pair of parallel spins in one unit cell\cite{Liu2017,Suzuki2017}) can have two chiralities. If the prior rotation was from $x$ to $y$, then  the wall will correspond to a clockwise rotation of spins ($+\pi/2$ chirality). On the other hand, when the prior rotation is from $-x$ to $y$, it will correspond to an anti-clockwise rotation of spins ( $-\pi/2$ chirality). In Fig.\ref{fig:Ratio} f) and g), each arrow represents the effective moment of each cluster octupole~\cite{Suzuki2017,Sugimoto2020}. Since the moment rotates from the center to the periphery the domain wall in the middle has a net moment perpendicular to the moment of the two domains sandwiching it.

\color{black}
\color{black}

The  amplitude of PHE depends on the strength of the prior magnetic field, as detailed in  Fig.\ref{fig:field dependent}. The raw data is shown in panel a). Similar to the case of Mn$_3$Sn~\cite{Li2019}, the larger the prior magnetic field, the lower the magnitude of PHE. Moreover, there is a difference between the evolution of the symmetric and the asymmetric  components of the PHE.  The asymmetric component, which is odd in magnetic field does not depend on the magnitude of the prior magnetic field (see panel c) ). The symmetric component, the genuine PHE, which is even in magnetic field  decreases as the the prior field is increases from 0.2 T to 0.8 T (panel c)).  Its amplitude decreases roughly proportional to the inverse of the prior field (panel (d)). This is compatible with a scenario in which the population of minority domains surviving after a field cycle determine the dominant chirality of domain walls. The larger the field, the lower the population of surviving minority domains, and the smaller the amplitude of the PHE signal. We found that the prior magnetic field required to suppress the PHE in Mn$_3$Ge is much smaller than that in Mn$_3$Sn at same temperature. This is compatible with the fact that the hysteresis loop is wider and the coercive fields larger in Mn$_3$Sn.

Now we turn our attention to the memory erasure by temperature cycling. The N\'{e}el temperature in Mn$_3$Sn is as high as 420 K, beyond what can be accessed to by a popular commercial instrument , namely Quantum Design's PPMS.  On the other hand, the N\'{e}el temperature in Mn$_3$Ge is 370 K and accessible. This allowed us to check that the PHE is wiped out when the sample is warmed above the ordering temperature.

Two measurement procedures were employed to obtain the data presented in Fig.\ref{fig:evolution} a,b). They are defined in the top panels of the figure. In the first procedure, dubbed field temperature cycling (FTC),  the sample is warmed up to  $T_\mathrm{cycl.}$ from   300 K keeping $\mu_{0}H = 0.2$T along $y$ and then cooled down to back to 300 K and then  PHE was measured by sweeping the field from -0.2 T to +0.2 T and back to -0.2 T . In the zero field temperature cycling (ZFTC) process,  the magnetic field was set to zero during the warm-up  to $T_\mathrm{cycl.}$ and cool-down to  300 K. Afterwards a PHE measurement was performed by sweeping the magnetic field in an identical fashion to the FTC process.

The PHE data obtained following these two ways of temperature cycling is displayed in Fig.\ref{fig:evolution}. The left panels show the FTC process and the right panels show the ZFTC process with several selected cycling temperatures. All measurements were performed at 300 K.  Fig.\ref{fig:evolution}b,f) show the PHE at the $T_\mathrm{meas.}$ = 300 K, peaking at 0.2 $\mu\Omega$ cm. When the cycling temperature is up to $T_\mathrm{cycl.}$ = 360 K, the peak value of residual PHE decreases to 0.1 $\mu\Omega$ cm in the FTC(see Fig.\ref{fig:evolution}c)) while it is 0.18 $\mu\Omega$ cm in the ZFTC (Fig.\ref{fig:evolution}g)). With further increasing $T_\mathrm{cycl.}$ to 370 K which is close to the $T_\mathrm{N}$ and cooling back to $T_\mathrm{meas.}$, only a slight residual PHE remains in Fig.\ref{fig:evolution}d). Remarkably, the PHE is barely changed in the ZFTC in the Fig.\ref{fig:evolution}h) while the cycling temperature approaches $T_\mathrm{N}$. The PHE disappears when $T_\mathrm{cycl}$ is higher than $T_\mathrm{N}$ in the FTC. For comparison, it can sustain beyond the $T_\mathrm{N}$ for ZFTC process, and it still has a finite value when $T_\mathrm{cycl.}$ = 380 K, slightly, but significantly, larger than  $T_\mathrm{N}$(see Fig.\ref{fig:evolution}i)).

The evolution of the amplitude of the PHE measured at $T$ = 300 K after  thermal cycling is shown as a function of $T_\mathrm{cycl.}$ in Fig.\ref{fig:compare}a). The temperature dependence of magnetization is also shown. Within experimental margin, the magnitude of field-cycled PHE  and $M (T)$ curve have a similar evolution. On the other hand, the amplitude of PHE after a ZFTC process is more robust and survives even when $T_\mathrm{cycl.}> T_\mathrm{N}$. This indicates that thermal fluctuations above $T_\mathrm{N}$ are strong enough to play a role in smearing out the chirality of magnetic domain  walls, Indeed, warming beyond $T_\mathrm{N}$ is expected to destroy the antiferromagnetic structure.

The temperature dependence of AHE and the peak value of the symmetric component  are compared in Fig.\ref{fig:compare}b).  Both  show a broad peak around 150 K, and become vanishingly small at the N\'{e}el temperature T$_N$ of 370 K.  The anomalous Hall conductivity, which can be extracted by combining Hall and residual resistivity is almost flat  at low temperature~\cite{Xu2020}. The temperature dependence of PHE is some what different. The evolution of the ratio of PHE/AHE with temperature is shown in Fig.\ref{fig:Ratio}(d)). It  steadily increases with decreasing temperature. Its peak amplitude is about 0.2, which is half of what was observed in Mn$_3$Sn\cite{Li2019}. This is presumably related to the fact that the hysteresis loop is wider and the coercive fields larger in the tin-based compound.

The manipulation of this PHE and memory effect by field and temperature cycling could be used in data storage. The PHE has two states (positive and negative) at room temperature, depending on the field history. By heating the device slightly higher than $T_\mathrm{N}$, can erase the direction information or binary information. So it could act as information store devices, realizing writing and erasing operations, while the AHE only depending on field orientation cannot fulfill the task. The rotating field process can be created by two perpendicular fields produced by two small coils in proximity to the device, so actual rotation of device or field is avoided. However, the coercive field in the Mn$_3$Ge is still large at 15 mT and the $T_\mathrm{N}$ are still high. If doping or other methods can reduces the $T_\mathrm{N}$ and coercive field, the device could be realized.

In summary, we observed the planar Hall effect in non-collinear antiferromagnetic Kagome Mn$_3$Ge and its memory effect down to 2 K. The ratio of PHE/AHE and the prior field to suppress the PHE are smaller than those in Mn$_3$Sn. Moreover, we erase this memory effect by temperature cycling through its N\'{e}el temperature with or without field, and find the sustain of PHE even the cycling temperature is higher than $T_\mathrm{N}$. This memory effect and memory erasing in PHE will be a new potential application in data storage device.

\textbf{Acknowledgements-} This work was supported by the National Key Research and Development Program of China (Grant no. 2016YFA0401704) and the National Science Foundation of China (Grant nos.51861135104 and 11574097).  In France, it was supported by the Agence Nationale de la Recherche  (ANR-18-CE92-0020-01; ANR-19-CE30-0014-04) and by Jeunes Equipes de l$'$Institut de Physique du Coll\`ege de France.  L. X. acknowledges a PhD scholarship by the China Scholarship Council(CSC).

\textbf{Date Availability-} The data that support the findings of this study are available from the corresponding author upon reasonable request

\noindent

* \verb|zengwei.zhu@hust.edu.cn|\\


\bibliography{ref}
\appendix

\end{document}